\documentclass{desyproc}

\usepackage[dvips]{hyperref}

\begin{document}

\title{Axions and other (Super-)WISPs\footnote{Talk presented at the {\it 5th Patras Workshop, Durham, UK, 13-17 July 2009}.}}

\author{{\slshape Markus Ahlers\footnote{Now at the C.N.~Yang Institute for Theoretical Physics, SUNY, Stony Brook, NY 11794-3840, USA}}\\[1ex]
Rudolf Peierls Centre for Theoretical Physics, University of Oxford, Oxford OX1 3NP, UK}

\contribID{ahlers\_markus}

\desyproc{DESY-PROC-2009-05}
\acronym{Patras 2009}
\doi  

\maketitle

\begin{abstract} 
We present some bottom-up motivations of axions and other weakly interacting sub-eV particles (WISPs) coupling to photons. Typically, these light particles are strongly constrained by their production or interaction in astrophysical and cosmological environments. Dedicated laboratory searches can provide complementary probes that are mostly less sensitive but also less model-dependent. We briefly comment on future experiments with the potential to discover photon oscillation effects in kinetic mixing scenarios with massive hidden photons. 
\end{abstract}

\section{Motivation}

Many extensions of the standard model predict hidden sectors of particles that are only weakly interacting with known matter. Some of these particles may even be extremely light (sub-eV), {\it e.g.}, if they are (pseudo-)Goldstone bosons of spontaneously broken (anomalous) global symmetries or gauge bosons of exact hidden symmetries. In some cases these light particles can be motivated by short-comings of the standard model, for instance the axion as a dynamical solution to the strong $\mathcal{CP}$ problem. We will give a brief outline of axion models in Sect.~\ref{sec:axion}. Other weakly interacting sub-eV particles\footnote{Axions can be considered as {\it super-WISPs} with a coupling inversely proportional to the Peccei-Quinn scale $F_a \gtrsim 10^9$~GeV (see Sect.~\ref{sec:axion}).} (WISPs) can be considered from their phenomenological point of view, having strong influence on early universe physics, astrophysics and even laboratory experiments despite their feeble interactions. In many cases these probes are independent of the particular origin of the light hidden sector and provide very general and simple test scenarios, {\it e.g.}, if the possible interaction of WISPs with standard model matter can be constrained by gauged and global symmetries. As an example, Sect.~\ref{sec:wisp} discusses mini-charged particles and hidden photons, that may naturally arise in field or string theoretic set-ups with hidden abelian gauge groups and kinetic mixing with the electromagnetic sector. We conclude in Sect.~\ref{sec:summary}. 

\section{Axions and their Relatives}\label{sec:axion}

Non-abelian gauge theories possess non-trivial solutions of the classical equations of motion in 4-dimensional euclidean space-time, so-called \emph{instantons}, that can be classified by an integer number,
the Pontryagin index
\begin{equation}
 q \equiv \frac{\alpha}{4\pi}\int{\rm d}^4 x\,{\rm tr}\,G_{\mu\nu}\widetilde G^{\mu\nu}\in \mathbf{Z}\,,
\end{equation}
where $G_{\mu\nu}$ is the field tensor of the non-abelian field with coupling $\alpha$ (see {\it e.g.} the reviews~\cite{CPreview}). 
Each instanton solution with $q=n$ is associated with a vacuum $|n\rangle$, that is left invariant under infinitesimal gauge transformations. However, there also exist gauge transformations with non-trivial winding number mapping between instanton solutions with different topological index. The true vacuum of non-abelian gauge theories is therefore a superposition of the vacua $|n\rangle$, the $\theta$-\emph{vacuum} \mbox{$\sum_{n\in\mathbf{Z}} \exp({\rm i}n\theta)|n\rangle$}.
The phase $\exp({\rm i}n\theta)$ contributes as an effective Lagrangian of the field theory,
\begin{equation}
\mathcal{L}_\theta = \theta\frac{\alpha}{4\pi}{\rm tr}\,G_{\mu\nu}\widetilde G^{\mu\nu}\,,
\end{equation}
which transforms as a pseudo-scalar and hence violates $\mathcal{CP}$.

The $\theta$-term of strong interaction is in general not invariant under chiral transformations in the presence of weak interactions and massive chiral fermions. The physical parameter is the combination $\bar\theta = \theta + \arg\det M$, which contributes to the neutron's electric dipole moment as $d_n\simeq 4.5\times10^{-15} \bar\theta\,e\,{\rm cm}$. The current limit of $|d_n| < 2.9\times10^{-26} e\,{\rm cm}$ translates into a limit of $|\bar\theta| \lesssim 10^{-10}$~\cite{Baker:2006ts,CPreview}. The \emph{strong $\mathcal{CP}$ problem} can now be formulated as the question why the sum of \emph{a priori} independent phases in $\bar\theta$ contributing to strong $\mathcal{CP}$-violation cancel with such a high accuracy.

An elegant solution to this problem has been proposed by \emph{Peccei} and \emph{Quinn}~\cite{PQ}. They introduced an anomalous global chiral symmetry U$(1)_{\rm PQ}$ which is spontaneously broken at the Peccei-Quinn (PQ) scale $f_a$. The axion corresponds to the pseudo-Goldstone boson of the broken symmetry, that receives a periodic potential at the quantum level~\cite{WW} due to chiral anomalies\footnote{For simplicity, we only consider QCD ($\propto\mathcal{N}$) and QED ($\propto\mathcal{E}$) contributions in Eq.~(\ref{eq:PQlag}).}, 
\begin{equation}\label{eq:PQlag}
V(a) = -\left(\bar\theta+\mathcal{N}\frac{a}{f_a}\right)\frac{\alpha_s}{4\pi}\,{\rm tr}\,G_{\mu\nu}\widetilde G^{\mu\nu}-\mathcal{E}\frac{a}{f_a}\frac{\alpha_{\rm em}}{8\pi}\,F_{\mu\nu}\widetilde F^{\mu\nu}\,.
\end{equation}
One can show that the QCD contributions ($\propto\mathcal{N}$) are bounded as $V(0)\leq V(a')$ in terms of a shifted axion field $a'\equiv a-\bar\theta F_a$ with effective PQ scale $F_a \equiv f_a/\mathcal{N}$. The strong $\mathcal{CP}$ problem is hence solved dynamically when the axion field settles down at its minimum $a=\bar\theta F_a$.

The original PQWW model~\cite{PQ,WW} includes a second Higgs doublet in the breaking of U$(1)_{\rm PQ}$ which relates the PQ breaking scale to the electroweak scale $F_a\simeq 246$~GeV. This model is now ruled out {\it e.g.}~by life-time measurements of mesons~\cite{CPreview}. Still viable variants of this model, so-called \emph{invisible} axion models, introduce additional Higgs bosons as electro-weak scalars which decouple the PQ scale from the weak scale. The mass of the axion can be determinant via current algebra techniques and is related to the ratio $z\equiv m_u/m_d \simeq0.35\div 0.6$ of up and down quarks together with the pion's mass $m_\pi$ and decay constant $f_\pi$ as
\begin{equation}\label{eq:axionmass}
m_a \simeq \frac{f_\pi m_\pi}{F_a}\frac{\sqrt{z}}{1+z}\simeq 6\,{\rm meV}\left(\frac{10^{9}\,{\rm GeV}}{F_a}\right)\,.
\end{equation}
The QED contribution ($\propto\mathcal{E}$) in Eq.~(\ref{eq:PQlag}) corresponds to a coupling term between axions and photons of the form $\mathcal{L}_{a\gamma\gamma} = -(g_{a\gamma\gamma}/4)a F^{\mu\nu}\widetilde F_{\mu\nu} = g_{a\gamma\gamma}a\,\mathrm{E}\cdot\mathrm{B}$ with
\begin{equation}\label{eq:axioncoupling}
g_{a\gamma\gamma} = \frac{\alpha_{\rm em}}{2\pi F_a}\left[ \frac{2}{3}\frac{4+z}{1+z} -\frac{\mathcal{E}}{\mathcal{N}}\right]\,.
\end{equation}
Note that the first term in the coupling~(\ref{eq:axioncoupling}) is a contribution form chiral symmetry breaking~\cite{CPreview}. The most popular examples are the DFSZ(-type) models~\cite{DFSZ} with $\mathcal{E}/\mathcal{N} = 8/3$ and KSVZ(-type) models~\cite{KSVZ} with $\mathcal{E}/\mathcal{N} = 0$.

Invisible axion models are constrained by their (model-dependent) axion coupling to matter and photons. In particular, astrophysical environments provide strong limits on the viable range of axion mass $m_a$ and coupling $1/F_a$ via photon conversion into axions in Compton-like scattering, by the Primakoff process or via hadronic or electromagnetic axion bremsstrahlung (for a review see~\cite{Raffelt:1996wa}). Depending on their production mechanism in the early universe, axions may also contribute today as (a part of) cold or warm dark matter and could be detected in tunable microwave cavitys~\cite{Sikivie:1983ip}. All these probes leave a window of viable axion models in the range $10^9\lesssim f_a/{\rm GeV}\lesssim 10^{12}$ and $10^{-5}\lesssim m_a/{\rm eV}\lesssim 10^{-2}$.

Generalizations of the QCD axion are \emph{axion-like particles} (ALPs). This class of particles includes pseudo-scalars $\phi$ with a photon interaction of type~(\ref{eq:axioncoupling}) but with masses and couplings kept as independent parameters. Pseudo-scalar ALPs may also originate via a PQ-type mechanisms and are generic in many supersymmetric extensions of the standard model (see {\it e.g.}~Ref.~\cite{PATRASconlon}). Strong bounds on the coupling $g_{\phi\gamma\gamma}$ arise from the direct search of ALPs produced in the Sun via helioscopes~\cite{Helioscopes}. Laboratory bounds from optical and light-shining-through-a-wall (LSW) experiment are typically three orders of magnitude weaker, but somewhat more model-independent~\cite{Jaeckel:2006xm}. Scalar ALPs with a coupling $\mathcal{L}_{\phi\gamma\gamma} = -\frac{1}{4}g_{\phi\gamma\gamma}\phi F^{\mu\nu} F_{\mu\nu} = \frac{1}{2}g_{\phi\gamma\gamma}\phi\left(\mathrm{E}^2-\mathrm{B}^2\right)$ can not be motivated by a PQ-type mechanism, but generally share the bounds of pseudo-scalar ALP models. In addition, scalar ALPs are strongly constrained by their contribution to non-newtonian forces~\cite{Dupays:2006dp}.

\section{WISPs from Kinetic Mixing}\label{sec:wisp}

Extensions of the standard model, in particular supersymmetric and/or string theories predict a plethora of additional particles, some of which might be extremely light~\cite{PATRASconlon}. These hidden sectors may couple to the standard model via renormalizable interactions, {\it e.g.}, via gauge kinetic mixing, the Higgs portal or Yukawa-type couplings. Hence, these interactions are not expected to be suppressed by the mass scale of messenger sectors. We will briefly discuss the case of weakly interacting sub-eV particles (WISPs) arising from kinetic mixing.

Gauge bosons $X_\mu$ of a hidden sector U$(1)_X$, so-called \emph{hidden photons} or \emph{paraphotons}, can couple to photons $A_\mu$ via a mixing term~\cite{Holdom:1985ag}
\begin{equation}
\mathcal{L}_{\rm mix} = -\frac{1}{2}\chi\,F^{\mu\nu}X_{\mu\nu}\,,
\end{equation}
where $F_{\mu\nu}$ and $X_{\mu\nu}$ are the field tensors of U$(1)_{\rm em}$ and U$(1)_X$, respectively.
Kinetic mixing naturally arises in field theoretic extensions of the standard model, where hidden sector particles are simultaneously charged under both U$(1)$s. If kinetic mixing is absent at the tree level, 1-loop diagrams give contributions of the form~\cite{Holdom:1985ag}
\begin{equation}
\chi \sim \frac{\sqrt{\alpha_{\rm em}\alpha_X}}{4\pi}\ln\left(\frac{m'}{m}\right)\,,
\end{equation}
where $m$ and $m'$ are the masses of a (non-degenerate) pair of hidden sector particles in the loop. Typically, field or string theoretical predictions of $\chi$ are in the range from $10^{-16}$ to $10^{-2}$~\cite{Dienes:1996zr} (see also Ref.~\cite{Goodsell:2009xc} and references therein).

The full kinetic Lagrangian can be diagonalized via a shift $X_\mu \to X_\mu -\chi A_\mu$ and a re-definition of the fine-structure constant $\alpha_{\rm em}\to\alpha_{\rm em}/(1-\chi^2)$. The shift of the hidden photon field can have several effects. If the hidden U$(1)_X$ is unbroken, hidden sector matter with a hidden charge $Q_X$ receives a electromagnetic charge of the form $Q_{\rm em} = -\chi\sqrt{\alpha_X/\alpha_{\rm em}}Q_X$, which can be extremely small for $\chi\ll1$ and/or $\alpha_X\ll\alpha_{\rm em}$, resulting in \emph{mini-charged particles} (MCPs)~\cite{Holdom:1985ag}. If the hidden photon is massive via a Higgs or St\"uckelberg mechanism the diagonalization of the kinetic term results in off-diagonal elements in the mass matrix. These terms are responsible for vacuum oscillations between photons and hidden photons with a probability~\cite{Okun:1982xi}
\begin{equation}
P_{\gamma\to\gamma'} = 4\chi^2\sin^2\left(\frac{m_{\gamma'}^2\ell}{4\omega}\right).
\end{equation}

Similar to the previous case, MCPs and hidden photons have a rich phenomenology with strong bounds coming from astrophysical and cosmological environments~\cite{Davidson:2000hf,Goodsell:2009xc}. The strongest bounds on the charge of sub-keV MCPs come from energy loss arguments of horizontal branch stars or white dwarves giving $|Q_X|\lesssim2\times10^{-13}$. Massive hidden photons in the sub-eV range receive also strong bounds from solar production and distortions of the CMB. However, due to the mass-dependence $P_{\gamma\to\gamma'} \propto m_{\gamma'}^4$ in the presence of short baselines, $\ell\ll\omega/m_{\gamma'}^2$, future experiments have the potential to probe previously unconstrained hidden photon models (see {\it e.g.}~Ref.~\cite{Goodsell:2009xc}). This includes LSW experiments with optical lasers~\cite{Ahlers:2007qf,Ehret:2009sq} and microwave cavities~\cite{Jaeckel:2007ch} or hidden photon helioscopes with extended baselines~\cite{Gninenko:2008pz}.

\section{Summary}\label{sec:summary}

We have discussed bottom-up motivations of light hidden sector particles that are weakly interacting with photons, so-called WISPs. Axions are introduced as a solution to the strong $\mathcal{CP}$ problem via the Peccei-Quinn mechanism. Other WISPs, in particular axion-like particles, mini-charged particles or hidden photons may arise in field or string theoretic extensions of the standard model. These light hidden sectors interacting with photons have a rich phenomenology, testable by early universe, astrophysical environments and laboratory experiments.  

\section*{Acknowledgments}

The author would like to thank the organizers of the {\it 5th Patras Workshop on Axions, WIMPs and WISPs} for a stimulating conference. He also wishes to thank H.~Gies, J.~Jaeckel, J.~Redondo and A.~Ringwald for the fruitful collaboration and acknowledges support by STFC UK (PP/D00036X/1) and the EU Marie Curie Network ``UniverseNet'' (HPRN-CT-2006-035863).

\begin{footnotesize}

\end{footnotesize}

\end{document}